\documentclass[useAMS,usenatbib]{mn2e}
\usepackage{graphics}
\usepackage{epsfig}
\usepackage{graphicx}
\usepackage[fleqn]{amsmath}

\newcommand{\mbh}{M_{\rm bh}}
\def\ltsima{$\; \buildrel < \over \sim \;$}
\def\simlt{\lower.5ex\hbox{\ltsima}}
\def\gtsima{$\; \buildrel > \over \sim \;$}
\def\simgt{\lower.5ex\hbox{\gtsima}}

\newcommand\ledd{{L}_{\rm Edd}}

\def\msun{{\,{\rm M}_\odot}}

\def\del#1{{}}

\title[IC signature of AGN feedback]{Inverse Compton X-ray
  signature of AGN feedback}

\author[]{Martin A. Bourne$^1$ and Sergei Nayakshin\\
Department of Physics \& Astronomy, University of Leicester, Leicester, LE1 7RH, UK\\
$^1$ {E-mail:~} {\rm martin.bourne@le.ac.uk}
}

\begin{document}

\date{Received}
\pagerange{\pageref{firstpage}--\pageref{lastpage}} \pubyear{2013}
\maketitle
\label{firstpage}

\maketitle

\begin{abstract}
Bright AGN frequently show ultra-fast outflows (UFOs) with outflow velocities
$v_{\rm out} \sim 0.1 c$. These outflows may be the source of AGN feedback on
their host galaxies sought by galaxy formation modellers. The exact effect of
the outflows on the ambient galaxy gas strongly depends on whether the shocked
UFOs cool rapidly or not. This in turn depends on whether the shocked
electrons share the same temperature as ions (one temperature regime; 1T) or
decouple (2T), as has been recently suggested. Here we calculate the Inverse
Compton spectrum emitted by such shocks, finding a broad feature potentially
detectable either in mid-to-high energy X-rays (1T case) or only in the soft
X-rays (2T). We argue that current observations of AGN do not seem to show
evidence for the 1T component. The limits on the 2T emission are far weaker,
and in fact it is possible that the observed soft X-ray excess of AGN is
  partially or fully due to the 2T shock emission. This suggests that UFOs
are in the energy-driven regime outside the central few pc, and must pump
considerable amounts of not only momentum but also energy into the ambient
gas. We encourage X-ray observers to look for the Inverse Compton components
calculated here in order to constrain AGN feedback models further.
\end{abstract}

\section{Introduction}

Super-massive black holes (SMBH) produce powerful winds
\citep{Shakura73,King03} when accreting gas at rates comparable to the
Eddington accretion rate. Such winds are consistent with the ``ultra-fast''
outflows (UFOs) $v_{\rm out} \sim 0.1 c$ detected via X-ray line absorption
\citep[e.g.,][]{PoundsEtal03b,PoundsEtal03a} and also recently in emission
\citep{PoundsVaughan11a}. The outflows must be wide-angle to explain their
$\sim 40$~\% detection frequency \citep{TombesiEtal10,Tombesi2010ApJ}. UFOs
may carry enough energy to clear out significant fractions of {\em all} gas
from the parent galaxy \citep[e.g.,][]{King10b,ZK12a} when they shock and pass
their momentum and perhaps energy to kpc-scale neutral and ionized outflows
with outflow velocities of $\sim 1000$ km~s$^{-1}$ and mass outflow rates of
hundreds to thousands of $\msun$~yr$^{-1}$
\citep[e.g.,][]{FergulioEtal10,SturmEtal11b,RP11a,LiuEtal13a}.

Most previous models of UFO shocks assumed a one-temperature model (``1T''
hereafter) where the electron and proton temperatures in the flow are equal to
each other at all times, including after the shock. \cite{FQ12a} showed that
shocked UFOs are sufficiently hot and yet diffuse that electrons may be much
cooler than ions (``2T'' model hereafter). They found that for an outflow
velocity of $0.1 c$ and $L_{Edd}=10^{46}$erg s$^{-1}$, the ion temperature is
$2.4\times 10^{10}$K but the electron temperature reaches a maximum of only
$T_e \sim 3\times 10^{9}$K in the post-shock region. The 1T regime may however
still be appropriate if there are collective plasma physics effects that
couple the plasma species tighter \citep[e.g.,][]{Quataert98}. There is thus a
significant uncertainty in how UFOs from growing SMBH affect their hosts,
e.g., by energy or momentum \citep{King10b}.

Here we propose a direct observational test of the 1T and 2T UFO shock
scenarios. AGN spectra are dominated by thermal disc emission coming out in
the optical/UV spectral region. The shocked electron temperature in both scenarios
is rather high, e.g., $T_e \sim 10^9$~K (2T) to $T_e \simgt 10^{10}$~K
(1T). Inverse Compton scattering of the AGN disc photons on these electrons
produces either soft X-ray (2T Inverse Compton; 2TIC) or medium to hard X-ray
energy (1TIC) radiation. Provided that the shock occurs within the Inverse
Compton (IC) cooling radius, $R_{\rm IC}\sim 500$~pc~$M_8^{1/2} \sigma_{200}$
(where $M_8$ is the SMBH mass in units of $10^8\msun$ and $\sigma_{200}$ is
the velocity dispersion in the host in units of 200~km~s$^{-1}$)
\citep{ZK12b}, essentially {\em all} the kinetic energy of the outflow,
$L_{\rm k} = (v_{\rm out}/2c) \ledd \sim 0.05 \ledd$  for $v_{out}=0.1$c, should be radiated away.
We calculate this IC spectral component and find it somewhat below but
comparable to the observed X-ray emission for a typical
AGN. Significantly, the IC emission is likely to be steady-state and
unobscured by a cold ``molecular torus'', which, for the 1T case, is in contrast to typical AGN X-ray
spectra. We therefore make a tentative conclusion that current X-ray
observations of AGN are more consistent with the 2T picture. In view of
the crucial significance of this issue to models of SMBH-galaxy co-evolution,
we urge X-ray observers to search for the 1TIC and 2TIC emission components in
AGN spectra to constrain the models of AGN feedback further.

\section{Inverse Compton Feedback Component}\label{sec:icfc}

\subsection{General procedure to calculate the X-ray spectrum}

In what follows we assume that the UFO velocity is $v_{\rm out}$, the total mass loss rate
is given by $\dot M = \ledd/(c v_{\rm out})$ and that the gas is pure hydrogen
in the reverse shock and so $n_{e} = n_{p}$. Assuming the strong shock jump 
conditions, the shocked UFO temperature
immediately past the shock is given by
\begin{equation}
k_B T_{\rm sh}= \frac{3}{16} m_p v_{\rm out}^2\;,
\label{tsh1}
\end{equation}
while the density of the shocked gas is
\begin{equation}
\rho_{\rm sh}=4\times\rho_{out}={4\times\frac{\dot M}{4\pi R^{2}v_{\rm out}}}={\frac{L_{\rm
      Edd}}{\pi R^{2}cv_{\rm out}^{2}}},
\label{rho_sh1}
\end{equation}
where $\rho_{out}$ is the pre-shocked wind density and $\dot M$ is the
  mass outflow rate in the wind. The factor of 4 in the density above comes
  from the density jump in the strong shock limit \citep{King10b,FQ12a}. The
shock is optically thin for radii $R \simgt 4G\mbh/v_{\rm out}^2 = 2 \times
10^{-3}$~pc~$M_8$.


The dominant cooling mechanism of the shocked wind is Inverse Compton
  (IC) Scattering \citep{King03}\footnote{Note that at low gas temperatures,
    $T < 10^7$~K, Compton processes instead heat the gas up
    \citep{CiottiOstriker07a}.}. Soft photons produced by the AGN are
  up-scattered by the hot electrons of the shocked wind to higher energies
  (X-rays for the problem considered here). Given the input spectrum of the
  soft photons and the energy distribution (EED, $F(\gamma)$ below, where
  $\gamma$ is the dimensionless electron energy, $E/m_e c^2$) of the hot
  electrons in the shock, one can calculate the spectrum of the IC up-scattered
  photons.

Consider first the case when the electron energy losses due to IC process are
negligible compared with the adiabatic expansion energy losses of the shocked
gas. In the zeroth approximation, then, we have a monochromatic population of
photons with energy ${\rm E_{0}}$ and total luminosity $L_0$ being up-scattered
by a population of electrons with a fixed Lorentz factor $\gamma $. The
typical energy of the up-scattered photons is given as ${\rm
  E_{f}\approx(\gamma^{2}-1)E_{0}}$. The emitted luminosity of these
up-scattered photons is given by
\begin{equation}
L_{\rm IC}=L_{0}\left(\frac{E_{f}}{E_{0}}\right)\tau
\end{equation}
$\tau$ is the Thompson optical depth of the shell, $\tau = \kappa_{\rm es}\rho
\Delta R$, where $k_{\rm es}$ is the electron Thompson scattering opacity,
$\rho$ is the shocked gas density and $\Delta R$ is the shell's thickness. To
arrive at the total luminosity of the IC emission one needs to calculate
$\tau$ as a function of time for the expanding shell. In any event, since we
assumed that IC losses are small, $L_{\rm IC} \ll L_{\rm k} = (v_{\rm out}/2c)
\ledd$, the kinetic luminosity of the ultra-fast outflow. This regime
corresponds to the shock extending well beyond the cooling radius $R_{\rm
  IC}$.

Here we are interested in the opposite limit, e.g., when the contact
discontinuity radius is $R\simlt R_{\rm IC}$, so that IC energy losses are
rapid for the shocked electrons. In this case the luminosity of the IC
emission is set by the total kinetic energy input in the shock, so that
\begin{equation}
L_{\rm IC}=L_{\rm k}\;.
\label{lic}
\end{equation}
On the other hand, one cannot assume that electron distribution of the shocked
electrons is constant.

Below we calculate this cooling electron distribution and the resulting IC
spectrum in both 1T and 2T regimes. We take into account that the input soft
photon spectrum is not monochromatic but covers a range of energies and the
electron population also has a distribution in $\gamma $. The spectral
luminosity density, $L_{\rm E_{f}}$, of the up-scattered photons, assumed to be
completely dominated by the first scattering\footnote{Since the wind shock is optically thin each photon should scatter once before escaping the system.} is given by,
\citep{NagirnerPoutanen94}:
\begin{equation}
\frac{d L}{d E_{\rm f}} = cE_{\rm f}\int_{1}^{\infty}d\gamma\frac{dF(\gamma
  )}{d\gamma}\int_{0}^{\infty}dE_{\rm 0}\frac{dn_{\rm{0}}}{dE_{\rm
    0}}\frac{d\sigma (E_{\rm f}, E_{\rm 0},\gamma )}{dE_{\rm f}}
\label{SLD}
\end{equation}
where $d n_{\rm 0}/d E_0=(1/4\pi R^{2}cE_{\rm 0})(dL_{\rm E_{\rm 0}}/dE_{\rm
  0})$ is the differential input photon number density at the location of the
shock (radius $R$), and $d\sigma(E_{\rm f}, E_{\rm 0},\gamma )/dE_{\rm f}$ is
the angle-averaged IC scattering cross-section for a photon of energy $E_0$ to
scatter to energy $E_{\rm f}$ by interacting with an electron of energy
$\gamma$ \citep{NagirnerPoutanen94}. 

The overall process to calculate the IC spectrum is as follows; in
  sections 2.2 and 2.3 the EED of the shocked electrons, $F(\gamma )$, is
  calculated. This part of the calculation is independent of the soft input
  spectrum, as long as the up-scattered photons are much less energetic than
  the electrons that they interact with. In order to calculate the output
  spectrum, however, we need to introduce the soft photon spectrum
  explicitly. These are model dependent since the precise physics, geometry
  and emission mechanism of the AGN accretion flows remains a work in
  progress. We therefore try three different models for the soft photon
  continuum: a black-body spectrum with $k_{B}T=3$eV, the UV region (1-100eV)
  of a typical AGN spectrum taken from \citet{SazonovEtAl04} and the entire
  ($1-10^{6}$eV) AGN spectrum taken from \citet{SazonovEtAl04}. Finally, the
  integrals in equation \ref{SLD} are calculated numerically and the total
  IC luminosity is normalised using equation \ref{lic}.

\subsection{The electron energy distribution in the 2T regime}

In the 2T regime for the shock, \cite{FQ12a} show that, while cooling behind
the shock, the electrons spend a considerable amount of time at a ``temporary
equilibrium'' state with temperature $T_{\rm eq}\sim 2 \times 10^9$~K for
$v_{out}=0.1$c (see figure 2 \citet{FQ12a}). Here we
therefore make the approximation that in the 2T regime the electrons have a
thermal EED at temperature $T = T_{\rm eq}$, described by the
Maxwell-J\"{u}ttner distribution,
\begin{equation} 
\frac{dF(\gamma )}{d\gamma} = n(\gamma,\theta ) =\frac{\beta\gamma^{2}}{\theta
    K_{2}\left(\frac{1}{\theta}\right)}e^{\frac{-\gamma}{\theta}},
\label{EED}
\end{equation}
where $\theta = k_{\rm B}T/(m_{\rm e}c^2)$, is the dimensionless electron
temperature and $K_{2}$ is the modified Bessel function of the second kind.

\subsection{1T cooling cascade behind the shock}\label{sec:1T}

Now we turn to the 1T case, assuming that the electron and ion temperatures in
the shocked UFOs are equal to one another at all times. In this case, there is
no ``temporary equilibrium'' state; behind the shock the electron temperature
drops with time from $T=T_{\rm sh}$ according to the IC cooling rate. The
absolute minimum temperature to which the electrons will cool is given by the
Compton temperature of the AGN radiation field, which is found to be $T_{\rm
  IC} = 2\times 10^7$~K by \cite{SazonovEtAl04}. The cooling of the electrons
leads to an electron temperature distribution being set up behind the shock
\citep{King10b} which we calculate here.

The electron-electron thermalisation time scale is $\sim m_e/m_p$ times
shorter than the energy exchange time scale with protons
\citep{Stepney83}. One can also show that IC electron losses even in the 1T
regime are not sufficiently large compared with electron self-thermalisation
rate to lead to strong deviations from the thermal distribution for the
electrons \citep[cf. equation 5 in][]{NayakshinMelia98}. We therefore assume
that the electrons maintain a thermal distribution behind the shock at all
times as they cool from the shock temperature $T_{\rm sh}$ to $T_{\rm
  IC}$. Our goal should thus be to calculate how much time electrons
  spend at different temperatures as they cool; this will determine
  $F(\gamma)$ and the resulting IC spectrum.

The rate of cooling due to the IC process is
\begin{equation}
\left(\frac{d u}{dt}\right)_{\rm IC} = -\frac{4}{3}\sigma_{\rm T}cU_{\rm
  rad}\int_{1}^{\rm\infty}\left(\gamma^2-1\right) n(\gamma ,\theta )d\gamma\;.
\label{dudt1}
\end{equation}
The plasma specific
internal energy density, $u$, is the sum of the ion contribution, $ (3/2)\,
k_{\rm B} T$, and that for the electrons. For convenience of notations we
define $u = a_e(\theta) \theta m_{\rm e}c^2$, where
\begin{equation}
a_e\left(\theta\right) =\frac{3}{2} + \frac{\left<\gamma\right>-1}{\theta}
\label{ae}
\end{equation}
and $\left<\gamma\right> = \int_{1}^{\infty} \gamma n(\gamma ,\theta
)d\gamma$ is the average electron $\gamma$-factor. Clearly, $a_e = 3$ and $a_e
= 9/2$ in the non-relativistic and extreme relativistic electron regimes,
respectively.  Finally, $U_{\rm rad} = \ledd/(4\pi R^2 c)$ is the energy
density of the AGN radiation field. We neglect the contribution of stars to
$U_{\rm rad}$.

We also need to include the compressional heating behind the shock
front, so that 
\begin{equation}
\frac{d u}{dt} = \left(\frac{d u}{dt}\right)_{\rm IC} -P\frac{dV}{dt}\;,
\label{dudt2}
\end{equation}
where ${P=(\Gamma -1)\rho u}$ is the pressure of the gas, $\Gamma$ is the
adiabatic index and $V=1/\rho$ is the specific volume of the gas.  Assuming
that the flow velocity is much smaller than the sound speed behind the shock,
 the region can be considered almost isobaric\footnote{The time it takes a sound wave to travel across
  the shocked wind is much less than the time it takes the shock pattern to
  propagate the same distance and so any fluctuations in the pressure will
  very quickly be washed out, see \citet{Weaver77}} i.e. Pressure$\approx constant$. 
  One finds $-PdV/dt = (1-\Gamma) du/dt$, so that the electron temperature evolution
   is solved from
\begin{equation}
m_e c^2 \frac{d}{dt}\left(a_e(\theta) \theta\right) = \frac{1}{\Gamma} \left(\frac{d
  u}{dt}\right)_{\rm IC}
\label{dudt3}
\end{equation}
This equation is solved numerically in order to determine $\dot \theta =
d\theta/dt$. One can define the dimensionless function $G(\theta )$,
\begin{equation}
G\left(\theta\right) = \frac{1}{t_{\rm c}}\frac{\theta}{\dot{\theta}}\;,
\label{gtheta1}
\end{equation}
where $t_{\rm c} =m_{\rm e}c^2 /(\sigma_{\rm T}cU_{\rm rad})$, is a timescale
factor which happens to be the order of magnitude of the IC cooling time for
non-relativistic electrons. 

\begin{figure}
	\centering \includegraphics[width=0.4\textwidth]{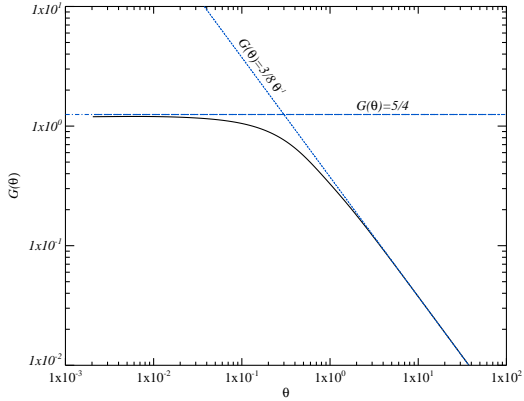}
	\caption{The dimensionless electron temperature distribution $G(\theta
          ) =\theta/(\dot{\theta}t_{\rm c})$. The dashed and dotted lines are labelled to show
          how the distribution behaves in the non-relativistic and
          ultra-relativistic regimes respectively. i.e. $G(\theta )=5/4$
            and $(3/8)\theta^{-1}$.}
	\label{thet_dist}
\end{figure}

We call $G(\theta)$ the Inverse Compton 1T cooling cascade (1TCC)
distribution, and plot it in Figure \ref{thet_dist}. Note that the
  function is independent from the outflow rate, $\dot M$, the energy density
  of the AGN radiation field, $U_{\rm rad}$, or the soft photon spectrum as
  long as the up-scattered photons are much less energetic than the electrons
  themselves. The function $G(\theta)$ is thus a basic property of the IC
  process itself.

We calculate $G(\theta)$ numerically and plot it in figure \ref{thet_dist}
below, but one can easily obtain the general form of the function in the two
opposite regimes analytically. \citet{RybickiLightman}
  show that in the non-relativistic (NR, $\theta << 1$) and ultra-relativistic
  (UR, $\theta >> 1$) limits the IC rate of cooling of a thermal distribution
  of electrons is given by
\begin{equation}
\frac{du}{dt}=-{c\sigma_{T}U_{rad}}\begin{cases}
4\theta \qquad \hbox{\rm non-relativistic } \\
16\theta^{2} \qquad \hbox{\rm ultra-relativistic}
\end{cases}
\end{equation}
Using these one can solve equation \ref{dudt3} analytically in the NR and UR limits to find:
\begin{equation}
\frac{d\theta}{dt}=-\frac{c\sigma_{T}U_{rad}}{m_{e}c^{2}} \begin{cases}
\frac{4}{3\Gamma} \quad \hbox{\rm    non-relativistic}\\
\frac{32}{9\Gamma}\theta^{2}\quad \hbox{\rm
    ultra-relativistic}
\end{cases}
\label{tsh2}
\end{equation}
and so $G(\theta )=5/4$ and $(3/8)\theta^{-1}$  in the NR and UR regimes
respectively.

The blue dashed and dotted lines in Figure \ref{thet_dist} show these limits,
highlighting that our solution for $G(\theta )$ behaves correctly in the
limiting regimes. The physical interpretation of the limiting forms of
$G(\theta)$ is quite clear. At high $\theta$, electrons are relativistic and
thus their IC cooling time is inversely proportional to $\theta$. Thus, the
hotter the electrons, the faster they cool. This yields the $G(\theta)\propto
\theta^{-1}$ behaviour at $\theta\gg 1$. In the opposite, non-relativistic
limit, the IC cooling time is independent of electron temperature, and this
results in $G(\theta )=5/4$ limit.

We now use $G(\theta )$ to calculate the ``integrated'' EED as seen by the soft AGN photons passing
through the shocked shell. The number of electrons with a temperature between
$\theta$ and $\theta + d\theta$ is given by $dN = (dN/d\theta) d\theta = \dot
N dt$, where $dt = d\theta/{\dot \theta}$ is the time that it takes electrons to cool from
temperature $\theta + d\theta$ to $\theta$, and $\dot N = \dot M/m_p$ is the
rate of hot electron ``production''. Clearly,
\begin{equation}
\frac{dN}{d\theta} = {\frac{\dot N}{\dot \theta}} =
\frac{\dot{N}t_{c}}{\theta}G(\theta )\;.
\label{dndt1}
\end{equation}

As electrons at each $\theta$ are distributed in the energy space according to
equation \ref{EED}, the number of electrons with $\gamma$-factor between
$\gamma$ and $\gamma+d\gamma$, $(d F(\gamma)/d\gamma) d \gamma$, is given by a
convolution of the thermal distribution $n(\gamma, \theta)$ with the electron
cooling history (function $dN/d\theta$):
\begin{equation}
\frac{d F(\gamma)}{d\gamma} = \int_{\theta_{\rm IC}}^{\theta_{\rm sh}} n(\gamma,
\theta) \frac{dN}{d\theta} d\theta\;,
\label{TAEED}
\end{equation}
where $\theta_{\rm sh} = k_{\rm B} T_{\rm sh}/(m_{\rm e} c^2)$, and
$\theta_{\rm IC} = k_{\rm B} T_{\rm IC}/(m_{\rm e} c^2)$.

\begin{figure}
	\centering
		\includegraphics[width=0.5\textwidth]{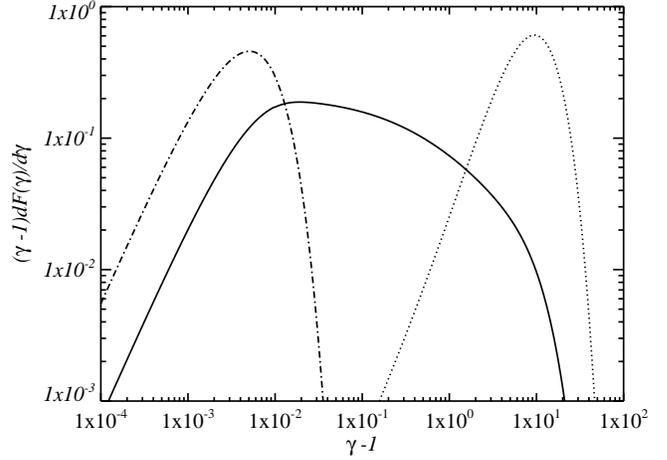}
	\caption{Electron energy distribution for 1T cooling cascade with
            $v_{\rm out}=0.1 c$ and $T_{\rm sh}=2\times 10^{10}$K (solid
            curve). For comparison, a single temperature thermal electron
            distributions are also shown for $T=T_{\rm sh}$ and $T=T_{\rm
              IC}=2\times 10^{7}$K with the dotted and dash-dotted curves,
            respectively.}
	\label{taeed}
\end{figure}

The cooling-convolved electron distribution function, $dF/d\gamma$, normalised
per electron in the flow, is shown in Figure \ref{taeed}. We assumed $v_{\rm
  out}=0.1 c$ and hence, $T_{\rm sh}=2\times 10^{10}$K. For comparison we also
plot the single temperature EEDs, $n(\gamma, \theta_{\rm sh})$, and $n(\gamma,
\theta_{\rm IC})$. This figure shows that in terms of number of electrons, the
distribution is strongly dominated by the lower-energy part, $\theta =
\theta_{\rm IC}$. This is because high energy electrons cool rapidly and then
``hang around'' at $T\approx T_{\rm IC}$. On the other hand, electron energy
losses are dominated by $\theta\approx \theta_{\rm sh}$ since these are
weighted by the additional factor $\sim (\gamma-1)^2$ (cf. equation
\ref{dudt1}). Since the EED is power-law like in a broad energy range, we
expect the resulting IC spectra to be power-law like in a broad range as well.

\section{Resulting spectra for 1T and 2T shocks}\label{sec:results}

\begin{figure}
	\centering
		\includegraphics[width=0.5\textwidth]{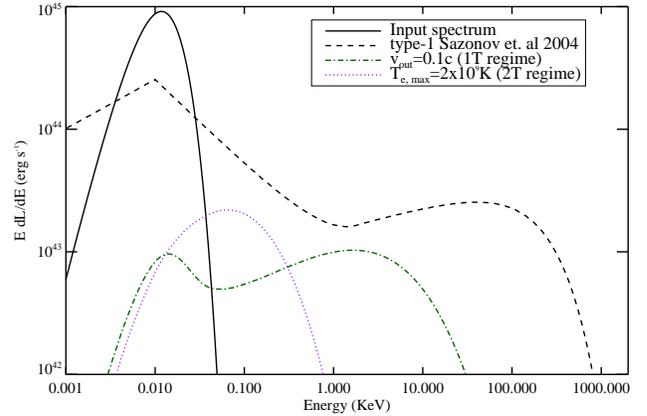}
	\caption{The Inverse Compton emission from shocked Ultra-Fast Outflows
          with $v_{\rm out} = 0.1 c$ in the 1T and 2T regimes (green dashed and purple dotted lines respectively). The primary soft photon spectrum from AGN, modelled as a
          simple black-body at $k_{B}T=3$eV, is also shown with a solid curve
          at low energies. The dashed curve shows a synthetic type 1 AGN spectrum from
          figure 4 of \citet{SazonovEtAl04}.}
          \label{fig:spectrum1}
\end{figure}

Figure \ref{fig:spectrum1} shows the Inverse Compton spectra in both the 2T
and 1T regimes, as labelled on the Figure. We assumed a SMBH of $\mbh = 10^7
\msun$ and outflow velocity of $v_{\rm out}= 0.1 c$. The input spectrum is
modelled by a black-body of single temperature $k_{B}T=3$~eV and bolometric
luminosity $L=L_{\rm bol}=\ledd$. This simple model assumes that the UV
luminosity of the innermost disc is absorbed and reprocessed into a cooler
black-body spectrum (we remind the reader that we assume that the UFO shocks
at ``large'' distances from the AGN, e.g., $R\sim 0.1 - 100$~pc). Also shown on
the plots, for comparison, is a synthetic spectrum of a type 1 AGN as computed
by \citet{SazonovEtAl04} normalised to the same bolometric luminosity. This
last spectral component demonstrates that both 1T and 2T spectral components
are actually comparable to the overall theoretical AGN spectra without UFOs; the 1T in the $\sim 2-10$
keV photon energy spectral window, whereas the 2T shock could be detectable in
softer X-rays.

\begin{figure}
	\centering
		\includegraphics[width=0.5\textwidth]{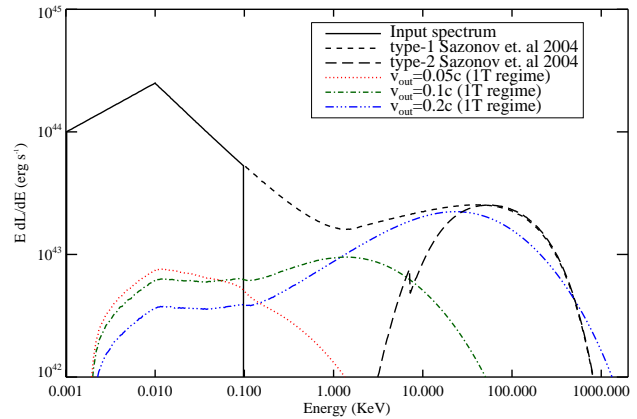}
	\caption{The Inverse Compton emission from shocked Ultra-Fast Outflows
          with $v_{\rm out} = 0.05$, $0.1$ and $0.2$c (red dotted, green
          dash-dotted and blue
          dash-double dotted respectively) in the 1T regime only. In contrast
          with fig. \ref{fig:spectrum1} the primary soft photon spectrum from the AGN is modelled by the the $1-100$eV region of the
          type 1 AGN spectrum from figure 4 \citet{SazonovEtAl04}. The black
          dashed and long-dashed curves shows synthetic type 1 and type 2 AGN spectra from
          figure 4 of \citet{SazonovEtAl04} respectively.}
	\label{fig:spectrum2}
\end{figure}

To explore the sensitivity of our results to model parameters, in Figure
\ref{fig:spectrum2} we use observationally-motivated soft photon spectra from
\citet{SazonovEtAl04} for energies below 0.1 keV, and we also consider two
additional values for the outflow velocity, $v_{\rm out}/c = 0.05$~and~$0.2$.
This figure also shows a synthetic type II (obscured AGN) spectrum from
\cite{SazonovEtAl04}, shown with the long-dash curve.

The figures demonstrate that at high enough outflow velocities, $v_{\rm out}
\sim 0.2 c$, the shocked UFOs produce power-law like spectra similar in their
general appearance to that of a typical AGN. In fact, we made no attempt to
fine tune any of the parameters of the \cite{King03} model to produce these
spectra, so it is quite surprising that they are at all similar to the
observed type I AGN spectra. In view of this fortuitous similarity of
  some of our IC spectra to the typical AGN X-ray spectra, one can enquire
  whether IC emission from $\sim pc$ scale shocks do actually contribute to
  the observed spectra.

Let us therefore compare the model predictions and X-ray AGN observations in
some more detail:
\begin{enumerate}

\item{{\em Bolometric luminosity.} Figures \ref{fig:spectrum1} and
  \ref{fig:spectrum2} are computed assuming 100\% conversion of the UFO's
  kinetic power in radiative luminosity, e.g., $L_{\rm IC} = L_{\rm k}$
  (cf. equation \ref{lic}) which is a fair assumption within the cooling
  radius, $R_{\rm IC}$, for the reverse shock \citep[which is $\sim$ hundreds
    of pc for the 1T and just a few pc for the 2T models, respectively,
    see][]{ZK12b,FQ12a}. The ratio between the X-rays and the soft photon
  radiation in our model is thus $\sim (v_{\rm out}/2c)$, e.g., 0.05 for
  $v_{\rm out} = 0.1 c$, which is just a factor of a few smaller than it is in
  the typical observed AGN spectra. In terms of shear bolometric luminosity
  1TIC and 2TIC are thus definitely observable.

When the shock front propagated farther than $R_{\rm IC}$, the overall
luminosity of the shock decreases. In the limit of extremely large
  $R_{\rm cd}$, where $R_{\rm cd}$ is the contact discontinuity radius, the
  primary outflow shocks at the radius $R_{\rm sw} \sim (1/5) R_{\rm cd}$
  \citep[see the text below equation 6 in][]{FQ12a}. When $R_{\rm sw} \simgt
  R_{\rm IC}$, the outflow is in the energy-conserving mode. We estimate that
  the IC luminosity would scale as $\propto R_{\rm IC}/R_{\rm sw}$ in this
  regime. In the intermediate regime, $ R_{\rm sw} < R_{\rm IC} < R_{\rm cd}$,
  $L_{\rm IC} < L_{\rm k}$.  A more detailed calculation is required in this
  regime to determine $L_{\rm IC}$ than has been performed here.

In the model of \cite{King03}, while the SMBH mass is below its critical
$M_\sigma$ mass, the outflows stall in the inner galaxy, $R\simlt R_{\rm
  ic}$. Once $\mbh > M_\sigma$, however, the outflow quickly reaches $R\sim
R_{\rm IC}$ and then switches over into the energy-conserving mode, which is
far more efficient. Therefore, we would expect that the 1TIC shock emission
should be a relatively widespread and relatively easily detectable feature in
this scenario. In the 2TIC case, however, $R_{\rm IC}$ is just a few
pc. Furthermore, since the outflow is much more likely to be in the energy
conserving mode, even SMBH below the $M_\sigma$ mass may clear galaxies. We
would expect shocks in this model spend most of the time in the regime $R_{\rm
  cd}\gg R_{\rm IC}$, being much dimmer than shown in figures
\ref{fig:spectrum1} and \ref{fig:spectrum2}. The 2TIC component is thus harder
to detect for these reasons.

\item{\em Variability.} The IC shocks are very optically thin, so that the
  observer sees an integrated emission from the whole spherical shocked
  shell. Accordingly, the IC shell emission cannot vary faster than on time
  scale of $ R_{\rm cd}/c \sim 30$~years~$R_{\rm cd}/(10$~pc). The shock
  travel time is even longer by the factor $c/v_{\rm out} \sim 10$. This
  therefore predicts that IC shock emission must be essentially a
  steady-state component in X-ray spectra of AGN. In contrast, observed X-ray
  spectra of AGN vary strongly on all sorts of time scales, from the duration
  of human history of X-ray observations, e.g., tens of years, to days and
  fractions of hour \citep[e.g.,][]{VaughanEtal03a}. This rapid
  variability is taken to be a direct evidence that observed X-rays must be
  emitted from very close in to the last stable orbit around SMBHs.

\item{\em No molecular torus obscuration in X-rays.} Nuclear emission of AGN,
  from optical/UV to X-rays, is partially absorbed in ``molecular torii''
  \citep{Antonucci93} of $\sim $~pc scale \citep{TristramEtal09}. This
  obscuration produces the very steep absorption trough in soft X-rays seen in
  type II AGN as compared with the type I sources (cf. long-dashed versus
  dashed curves in Fig.\ref{fig:spectrum2}). If a sizeable fraction of X-ray
  continuum from AGN were arising from the IC shocks on larger scales, then
  that emission would not show any signatures of nuclear X-ray
  absorption. While \cite{GalloEtal13} reports one such ``strange'' AGN, it is
  also a very rapidly varying one (cf. their figs. 9 and 10), which again rule
  out the 1TIC model. There are also examples when soft X-ray absorption
  varied strongly on short time scales \citep[e.g.,][]{PuccettiEtal07},
  indicating that X-ray emission region is as small as $10^{-4}$ pc.

\item{\em No reflection component.} Compton down scattering and soft-X-ray
  absorption by circum-nuclear gas produces the reflection component or
  ``Compton hump'' observed in many AGN at $\sim30$ KeV
  \citep{GuilbertRees88,Pounds90}.  In addition, the fluorescent Fe
  K-$\alpha$ line emission is associated with the same process and is
  frequently detected in X-ray spectra of AG \citep{NandraPounds94}. Since the
  shocks that we study occur on large scales, the IC emission would likely
  impact optically thin cold gas and thus result in much weaker X-ray
  reflection and Fe K-$\alpha$ line emission than actually observed.
}

\end{enumerate}

\begin{figure}
	\centering
		\includegraphics[width=0.5\textwidth]{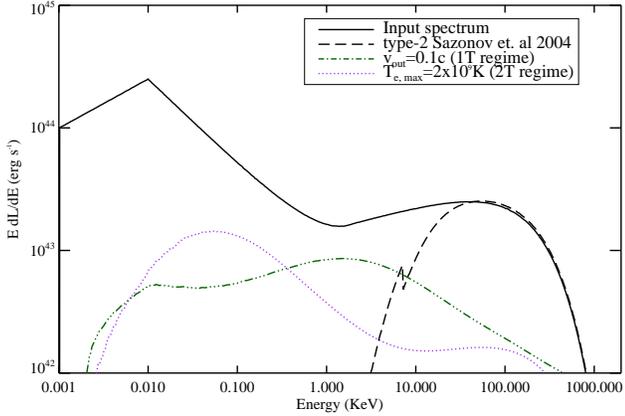}
	\caption{Same as figure \ref{fig:spectrum1}, but now assuming that the
          AGN spectrum in exactly equal to the model of
          \citet{SazonovEtAl04}. Again we show the case where $v_{\rm out} =
          0.1 c$  in the 1T and 2T regimes (green dashed and purple dotted lines respectively).}
	\label{fig:spectrum3}
\end{figure}

Given these points, we can completely rule out the most extreme assumption
  that the X-ray emission of AGN is due to UFO shocks alone. The next question
  to ask is whether having the 1TIC or 2TIC emission from the UFOs in {\em
    addition} to the ``nuclear'' X-ray corona emission of AGN \citep{Haardt93}
  would be consistent with the present data. To address this, we calculate the
  1TIC and 2TIC emission as for figure \ref{fig:spectrum2}, but now including
  the part of the \cite{SazonovEtAl04} spectrum above 0.1 keV, which means
  that we now also include IC scattering of the higher energy radiation from
  AGN in the UFO shocks (rather than only the disc emission).  The resulting
  spectra are shown in Figure \ref{fig:spectrum3}.

We see that the 1TIC spectra would be ruled out in deeply absorbed type II AGN
spectra, because the 1TIC component would be very obvious in these sources
below a few keV. The 2TIC component, on the other hand, would not be so
prominent except in very soft X-rays where interstellar absorption is
significant. We therefore preliminarily suggest that X-ray emission from 1T UFO
shocks may contradict the data for type II AGN, whereas 2TIC spectra would
probably be comfortably within the observational limits. 

\section{Discussion and Conclusions}\label{sec:discussion}

We calculated X-ray spectra of 1T and 2T Inverse Compton shocks resulting from
ultra-fast outflows from AGN colliding with the ambient host galaxy
medium. We concluded that 1TIC spectra could be detectable in AGN spectra
  and distinguisheable from ``typical'' AGN spectra actually observed by
  absence of rapid variability, Compton reflection and Fe K-$\alpha$
  lines. This disfavours 1T models for AGN feedback in our opinion.
We must nevertheless caution that the quoted typical observed AGN spectra and
properties may be dominated by local objects that are simply not bright enough
to produce a significant kinetic power in outflows,
which our model here assumed. We therefore urge X-ray observers to search for
the un-absorbed and quasi-steady emission components presented in our paper in
order to clarify the situation further.

It is interesting to note that the 2T Inverse Compton emission (2TIC)
  comes out mainly in the soft X-rays where it is far less conspicuous as this
  region is usually strongly absorbed by a cold intervening absorber. In fact,
  it is possible that 2TIC emission component calculated here does contribute
  to the observed ``soft X-ray excess'' feature found at softer X-ray energies
  ($< 1$ KeV) but not yet understood
  \citep{GierliDone04,RossFabian05,Crummy06,Scott12}.  The 2T spectral
  component in figure \ref{fig:spectrum1} is close to the observed shape of
  the soft excess and would provide a soft excess that is independent of the
  X-ray continuum, a requirement suggested by e.g. \citet{Rivers12}.  The
  observed soft excess does not vary in spectral position over a large range
  of AGN luminosities \citep{WalterFink93, GierliDone04, Porquet04}. The 2TIC
  model may account for this as well since figure 2. of \citet{FQ12a} shows
  that $T_{\rm {eq}}$ is quite insensitive to the exact value of the outflow
  velocity. Finally, the 2TIC emission would exhibit little time variability.
  \cite{UttleyEtal03,PoundsVaughan11a} reported a quasi-constant soft X-ray
  component in NGC~4051 which could only be seen during periods of low (medium
  energy) X-ray flux, which is qualitatively consistent with the 2TIC shock
  scenario.

Therefore, we conclude that general facts from present X-ray
observations of AGN not only disfavour 1TIC component over 2TIC component but
may actually hint on the presence of a 2TIC one in the observed spectra.

Whether the electrons thermally decouple from hot protons is vitally important
for the problem of AGN feedback on their host galaxies. Because of their far
larger mass, the ions carry virtually all the kinetic energy of the
outflow. At the same time, ions are very inefficient in radiating their energy
away compared with the electrons. In the 1T model, the electrons are able to
sap away most of the shocked ions energy and therefore the AGN feedback is
radiative, that is, momentum-driven, inside the cooling radius
\citep{King03}. In this scenario only the momentum of the outflow affects the
host galaxy's gas.  In the 2T scenario, the outflow is non-radiative, so that
the ions retain most of their energy. The AGN feedback is thus even more
important for their host galaxies in this energy-driven regime
\citep{ZK12a,FQ12a}.

If the outflows are indeed in the 2T mode, then one immediate implication
concerns the recently discovered {\em positive} AGN feedback on their host
galaxies. Well resolved numerical simulations of \cite{NZ12,ZubovasEtal13a}
show that ambient gas, when compressed in the forward shock (to clarify, the
shock we studied here is the reverse one driven in the primary UFO), can cool
rapidly in the gas-rich host galaxies. The nearly isothermal outer shock is
gravitationally unstable and can form stars. In addition,
\cite{ZubovasEtal13b} argue that galactic gas discs can also be pressurised
very strongly by the AGN-driven bubble. In these cases AGN actually have a
positive -- accelerating -- influence on the star formation rate in the host
galaxy. Within the 1T formalism, the AGN-triggered starbursts occur outside
$R_{\rm IC}\sim$ hundreds of pc only \citep{ZubovasEtal13b}. If outflows are
2T then AGN can accelerate or trigger star bursts even in the nuclear regions
of their hosts.

\section*{Acknowledgments}

The authors thank the anonymous referee for comments that helped improve the
manuscript. Theoretical astrophysics research in Leicester is supported by an STFC
grant. MAB is supported by an STFC research studentship. We thank Ken Pounds,
Andrew King, Sergey Sazonov, and Kastytis Zubovas for extended discussions and
comments on the draft. Sergey Sazonov is further thanked for providing
\cite{SazonovEtAl04} spectra in a tabular form.

\bibliographystyle{mnras}
\bibliography{./nayakshin.bib}

\label{lastpage}

\end{document}